\documentclass[aps,prb,twocolumn,superscriptaddress,showpacs]{revtex4-1}
\usepackage{graphicx}

\begin{document}

\title{Structural contributions to the pressure-tuned charge-density-wave to superconductor transition in ZrTe$_{3}$: Raman scattering studies }

\author{S.~L.~Gleason}\thanks{These coauthors contributed equally to this work.}
\author{Y.~Gim}\thanks{These coauthors contributed equally to this work.}
\author{T.~Byrum}
\author{A.~Kogar}
\author{P.~Abbamonte}
\author{E.~Fradkin}
\author{G.~J.~MacDougall}
\author{D.~J.~Van Harlingen}
\affiliation{Department of Physics and Frederick Seitz Materials Research Laboratory, University of Illinois, Urbana, Illinois 61801, USA}
\author{Xiangde Zhu}\thanks{Present address: High Magnetic Field Laboratory, Chinese Academy of Sciences, Hefei 230031, PRC.}
\author{C.~Petrovic}
\affiliation{Condensed Matter Physics and Materials Science Department, Brookhaven National Laboratory, Upton, New York 11973, USA}
\author{S.~L.~Cooper}
\affiliation{Department of Physics and Frederick Seitz Materials Research Laboratory, University of Illinois, Urbana, Illinois 61801, USA}

\date{\today}

\begin{abstract}
Superconductivity evolves as functions of pressure or doping from charge-ordered phases in a variety of strongly correlated systems, suggesting that there may be universal characteristics associated with the competition between superconductivity and charge order in these materials.
We present an inelastic light (Raman) scattering study of the structural changes that precede the pressure-tuned charge-density-wave (CDW) to superconductor transition in one such system, ZrTe$_{3}$.
In certain phonon bands, we observe dramatic linewidth reductions that accompany CDW formation, indicating that these phonons couple strongly to the electronic degrees of freedom associated with the CDW.
The same phonon bands, which represent internal vibrations of ZrTe$_3$ prismatic chains, are suppressed at pressures above $\sim$10~kbar, indicating a loss of long-range order within the chains, specifically amongst intrachain Zr-Te bonds.
These results suggest a distinct structural mechanism for the observed pressure-induced suppression of CDW formation and provide insights into the origin of pressure-induced superconductivity in ZrTe$_3$.
\end{abstract}


\maketitle

\section{Introduction}
One of the most exciting areas of condensed matter research involves the study of how superconductivity evolves from magnetic- or charge-ordered phases in a diverse range of strongly correlated systems, including high $T_{c}$ cuprates,\cite{Kiv.2007} iron-arsenide superconductors,\cite{doi:10.1146/annurev-conmatphys-070909-104041} and charge density wave materials.\cite{PhysRevLett.103.236401, Mor2006, Bar2008, Joe2014}
While the presence of secondary ordered phases are generally inimical to superconductivity in conventional superconductors, there is growing evidence that the optimal conditions for superconductivity in some unconventional superconductors may require some phase competition, and perhaps even phase coexistence.\cite{Kiv.2007,Joe2014}

A particularly interesting group of materials in which phase competition between charge and superconducting order appears to be important is charge-density-wave (CDW) materials that can be tuned to superconductor phases with either pressure or intercalation.
Examples of these systems include intercalated \cite{Mor2006} and pressure-tuned \cite{PhysRevLett.103.236401} 1T-TiSe$_{2}$, intercalated TaS$_{2}$,\cite{PhysRevB.78.104520} and intercalated \cite{Lei2011,PhysRevLett.106.246404} and pressure-tuned \cite{PhysRevB.71.132508} ZrTe$_{3}$.
These systems offer the opportunity to explore the competition between charge-order and superconductivity,\cite{PhysRevLett.103.236401, Mor2006, Bar2008, Joe2014} and particularly the structural changes that enable the emergence of superconductivity as the CDW state collapses with increasing pressure or intercalation.

\begin{figure}[h]
\includegraphics{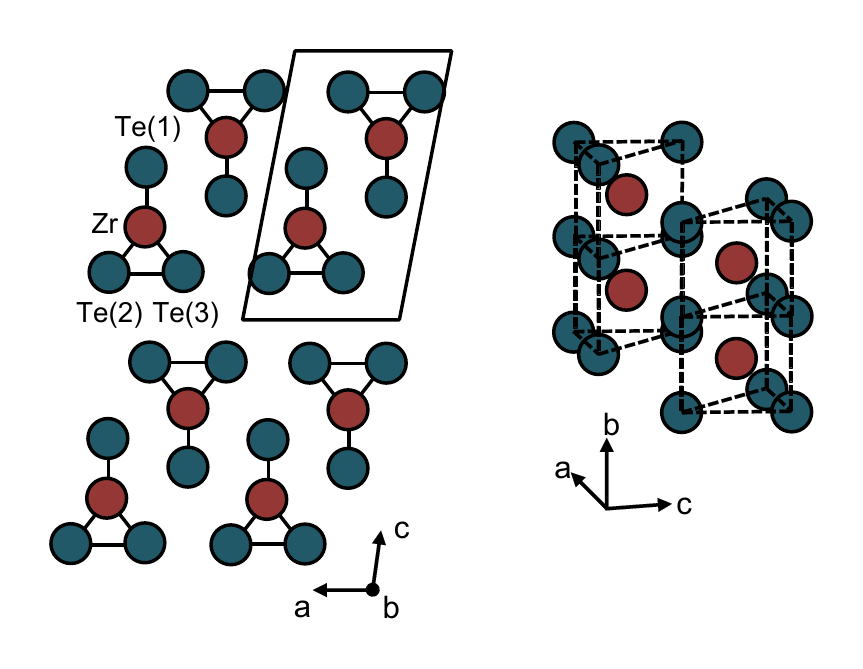}
\caption{\label{fig_lattice} (Color online) ZrTe$_{3}$ crystal structure.
(left) Projection onto the $ac$ plane.
(right) Perspective showing two ZrTe$_{3}$ prismatic chains.}
\end{figure}

ZrTe$_{3}$ crystallizes in the ZrSe$_{3}$-type structure (space group $P2_1/m$, shown in Fig.~\ref{fig_lattice}) in which ZrTe$_{3}$ trigonal prisms stack along the $b$ axis to form infinite quasi-one-dimensional chains.\cite{Sto1998}
The monoclinic unit cell contains two such chains, which are related by screw symmetry.
Interprism Zr-Te bonds hold the chains together to form layers in the $ab$-plane which are easily cleaved along the $c$ axis.
Within each layer, Te(2) and Te(3) atoms form a dimerized chain along the $a$ axis.\cite{Sto1998}

Electronically, ZrTe$_{3}$ exhibits an anisotropic metallic resistivity below 300 K $(\rho_{b}\sim \rho_{a}\sim \rho_{c}/10)$,\cite{PhysRevLett.106.246404, Tak1984} and has resistive anomalies associated with CDW formation at $T_{CDW} \approx 63$~K\cite{Tak1984} and filamentary superconductivity below $T_{c} = 2$~K.~\cite{Nak1986}
The CDW transition in ZrTe$_{3}$ is associated with a commensurate structural modulation of wavevector $\textbf{q}=(1/14,0,1/3)$, \cite{PhysRevLett.106.246404, Tak1984, Eag1984} and most strongly affects the conductivity perpendicular to the prismatic chains.\cite{Tak1984}
A Kohn anomaly associated with a soft phonon mode \cite{Hoe2009.2}---as well as CDW fluctuations that extend well above $T_{CDW}$\cite{Yok2005,Perucchi2005}---have been identified with the CDW transition in ZrTe$_{3}$.
Both band structure calculations \cite{Fel1998,Sto1998} and angle-resolved photoemission spectroscopy measurements\cite{Yok2005} indicate that the dimerized chains of Te(2)-Te(3) atoms along the $a$ axis yield quasi-1D electronlike sheets of Fermi surface, and that nesting of these sheets may be responsible for CDW formation.

The application of hydrostatic pressure increases $T_{CDW}$ with increasing pressure up to $\sim$10~kbar (1~GPa), but increasing the pressure above $\sim$20~kbar suppresses $T_{CDW}$, eventually leading to a complete suppression of CDW behavior above 50~kbar.\cite{PhysRevB.71.132508}
Increasing hydrostatic pressure also initially suppresses superconductivity for pressures greater than 5~kbar,\cite{PhysRevB.71.132508} but reentrant superconductivity is observed for pressures above 50~kbar, i.e., above the pressure at which CDW behavior is suppressed.\cite{PhysRevB.71.132508}
Cu- and Ni-intercalation have also been shown to suppress the charge density wave state and induce superconductivity in ZrTe$_{3}$.\cite{Lei2011, PhysRevLett.106.246404}

While much is known about the complex phase diagrams of pressure-tuned and intercalated ZrTe$_{3}$ systems, several key issues concerning the underlying mechanisms governing the CDW-to-superconductor transitions remain uncertain.
In particular, neither the relationship between the collapse of CDW order and the emergence of superconductivity in these systems, nor the nature of the structural changes that accompany this evolution, are well understood.
These unresolved issues---which also confront a much broader range of materials exhibiting CDW-to-superconductor transitions\cite{PhysRevLett.103.236401, Mor2006, Bar2008, Joe2014, PhysRevB.78.104520}---provide impetus for investigating the CDW-to-superconductor transition in ZrTe$_{3}$ using methods capable of probing the underlying structural changes associated with this transition.

We have used variable-temperature and variable-pressure inelastic light (Raman) scattering to investigate the microscopic details underlying the pressure-induced phases of ZrTe$_{3}$.
Phonon Raman scattering is particularly effective for studying the complex structural phases and changes in electron-phonon coupling that accompany temperature- and pressure-dependent phase changes in correlated materials such as ZrTe$_{3}$, because this technique can convey detailed information about changes associated with specific atomic elements of the unit cells.
In this paper, we show that certain phonon bands undergo dramatic linewidth reduction near $T_{CDW}$, indicating that these phonons couple strongly to the electronic degrees of freedom associated with the CDW.
The same phonon bands, which represent internal vibrations of the ZrTe$_3$ prismatic chains, are suppressed at pressures above $\sim$10~kbar.
This indicates a pressure-induced loss of long-range order within the chains, specifically amongst intrachain Zr-Te bonds.
We also find structural evidence for pressure-induced dimensional crossover in ZrTe$_{3}$.
These results suggest a distinct structural mechanism for the observed suppression of CDW formation above $\sim$20~kbar and provide insights into the origin of pressure-induced superconductivity in ZrTe$_3$.

\section{Experiment}

\subsection{Sample preparation}
Single crystals of ZrTe$_{3}$ were prepared at Brookhaven National Laboratory by chemical vapor transport, using a nearly stoichiometric mixture of powdered Zr and Te that was enclosed in an evacuated and sealed quartz ampoule along with iodine as the transport agent.\cite{Lei2011, PhysRevLett.106.246404}
The furnace gradient was kept between 760 and 650~C after heating to 700~C for two days.
Crystals were oriented using a Panalytical X'pert single-crystal x-ray diffractometer with Cu~K$_{\alpha 1}$ radiation.

\subsection{Raman scattering measurements}
Raman scattering measurements were performed using the 647.1~nm excitation line of a Kr$^{+}$ laser.
The incident laser power was limited to 5~mW and was focused to a $\sim$50~$\mu$m-diameter spot to minimize laser heating of the samples; consequently, it was assumed that there was a negligible increase in the sample temperature from laser heating.
The scattered light from the samples was collected in a backscattering geometry, dispersed through a triple-stage spectrometer, and then recorded with a liquid-nitrogen-cooled CCD detector.
The incident light polarization was selected with a polarization rotator, and the scattered light polarization was analyzed with a linear polarizer, providing symmetry information about the excitations studied.

Variable-pressure and variable-temperature measurements were performed using a miniature cryogenic diamond anvil cell inserted into a helium-flow-through cryostat.\cite{Sno2003}
This high pressure cell allows for \emph{in situ} pressure adjustment, enabling Raman scattering measurements at temperatures 3--300~K and pressures 0--100~kbar.
A liquid-argon medium was used to provide quasi-hydrostatic pressure.
The pressure was determined from the shift in the R$_1$ fluorescence line of a ruby placed near the sample.~\cite{Block1976}

\section{Results and discussion}

\subsection{Identification of phonon modes in ZrTe$_{3}$}

\begin{figure}[h]
\includegraphics{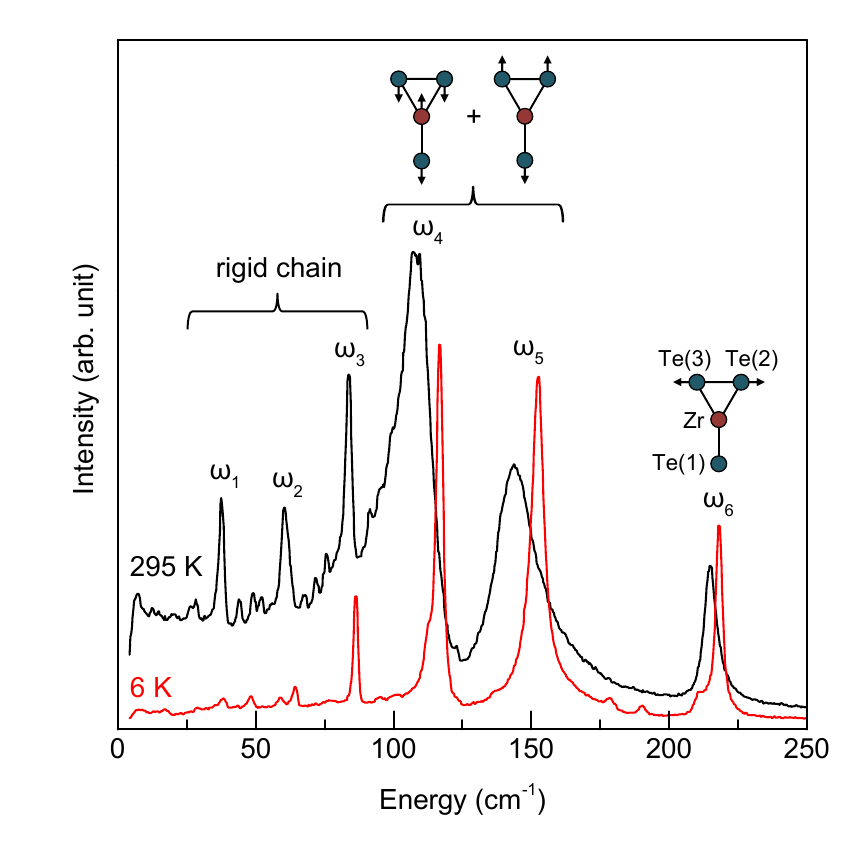}
\caption{\label{fig_mode_assignment} (Color online) Raman spectra of ZrTe$_{3}$ at $T=295$~K and $T=6$~K with normal mode displacement patterns.
Phonons $\omega_1-\omega_3$ involve vibrations of rigid ZrTe$_3$ chains against one another.
Phonons $\omega_4$ and $\omega_5$ involve deformations of the ZrTe$_3$ chains along the $c$ axis.
Phonon $\omega_6$ is a diatomic vibration.
Eigenmode illustrations after Ref.~\onlinecite{Zwi1979}.}
\end{figure}

Raman scattering has been used previously to measure the phonon spectrum of ZrTe$_3$ at room temperature\cite{Wieting1980} and at $T=77$~K.\cite{Zwi1980}
Fig.~\ref{fig_mode_assignment} shows our measurements of the phonon spectra of ZrTe$_{3}$ at $T=295$~K and 6~K, obtained in a $(\mathbf{E}_{i},\mathbf{E}_{s}) = (a,a)$ scattering geometry, where $\mathbf{E}_{i}$ and $\mathbf{E}_{s}$ are the incident and scattered light polarization directions, respectively.
The $T=295$~K phonon spectrum of ZrTe$_{3}$ exhibits six first-order phonon modes at $\omega_{1}=38$~cm$^{-1}$, $\omega_{2}=60$~cm$^{-1}$, $\omega_{3}=84$~cm$^{-1}$, $\omega_{4}=108$~cm$^{-1}$, $\omega_{5}=144$~cm$^{-1}$ and $\omega_{6}=215$~cm$^{-1}$.
These energies are similar to those obtained previously.\cite{Zwi1980,Wieting1980}

As mentioned, the ZrTe$_3$ crystal may be viewed as infinite chains of ZrTe$_3$ trigonal prisms that are held together by interprism Zr-Te bonds.
The phonon spectra of isostructural crystals ZrS$_3$ and ZrSe$_3$ are comprised of two sets of three vibrations that are well separated in energy\cite{Zwi1980}, encouraging a normal mode description that discriminates between `external' or prismatic-chain-preserving vibrations, and `internal' or prismatic-chain-distorting vibrations.
Although the two sets of vibrations are not as well separated in ZrTe$_3$ (see Fig.~\ref{fig_mode_assignment}), we nevertheless follow previous authors~\cite{Zwi1979,Wieting1980,Wieting1981} in assigning lower-energy modes $\omega_1$--$\omega_3$ to primarily external vibrations of the trigonal prismatic chains, and higher-energy modes $\omega_4$--$\omega_6$ to primarily internal vibrations of the trigonal prismatic chains.
These assignments are supported by the distinct pressure dependences of the external and internal mode frequencies (see section \ref{sec_pressure}), as well as the symmetry considerations described below.

In the limit of weak interchain bonding, such that the $2mm$ symmetry of the chains predominates, the ($a$,$a$) scattering geometry used in this experiment allows for the observation of only 3 A$_1$ internal vibrations.
One of these, a `diatomic' vibration of the Te(2)-Te(3) bond, is assigned to $\omega_6$ following previous lattice dynamical calculations.\cite{Wieting1981}
The other two internal vibrations, which involve deformations of the trigonal prisms along the $c$ axis, are assigned to $\omega_4$ and $\omega_5$.
As interchain bonding becomes more significant, the $2/m$ crystal symmetry allows for the observation of up to five additional vibrations (A$_1$ + 4 B$_2$) in the (a,a) scattering geometry, including three external vibrations of the prismatic chains.\cite{Fateley1972}
These chain preserving modes are assigned to the relatively weaker, lower energy modes $\omega_1$--$\omega_3$.

In addition to the sharp phonon modes shown in Fig.~\ref{fig_mode_assignment}, at $T=295$~K there is a broad background that is likely associated with inelastic electronic scattering.
This interpretation is supported by the broad and asymmetric Fano lineshapes\cite{Fan1866} of the $\omega_{4}$ and $\omega_{5}$ phonon modes, suggesting that these modes are particularly strongly coupled to the underlying electronic background.

Several weak features appear in the $T=6$~K spectrum, for example the peaks between $\omega=175$--200~cm$^{-1}$.
These features likely represent phonons that have been folded to $k=0$ from elsewhere in reciprocal space due to the CDW modulation of the structure.
As can be seen in the following sections, these peaks vanish when the system is tuned out of the CDW phase via either temperature or pressure.

\subsection{Temperature dependence of phonon modes in ZrTe$_{3}$}

\begin{figure}[h]
\includegraphics{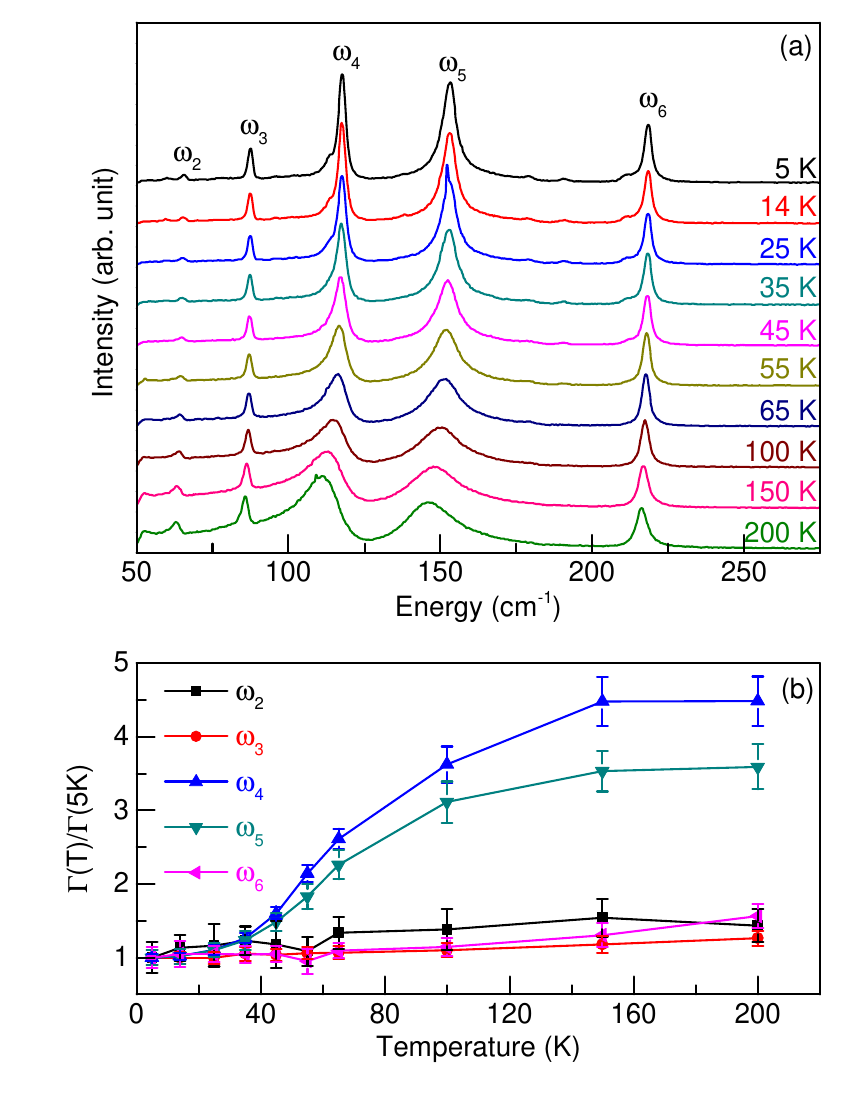}
\caption{\label{fig_temp} (Color online) (a) Temperature dependence of the ZrTe$_{3}$ Raman spectrum at ambient pressure.
Spectra have been offset for clarity.
(b) Phonon band linewidths, relative to $T=5$~K, as a function of temperature.}
\end{figure}

The temperature dependence of the Raman spectrum of ZrTe$_{3}$ is shown in Fig.~\ref{fig_temp}(a), with the temperature dependences of the phonon band linewidths summarized in Fig.~\ref{fig_temp}(b).
The rigid chain phonon modes $\omega_2$ and $\omega_3$, along with the diatomic mode $\omega_6$, show little linewidth dependence on decreasing temperature.
On the other hand, vibrations $\omega_{4}$ and $\omega_{5}$, which involve deformations of the ZrTe$_{3}$ trigonal prisms along the $c$ axis, exhibit dramatic changes in linewidth as a function of decreasing temperature.
In particular, between $T = 200$~K and 5~K the linewidths of the $\omega_{4}$ and $\omega_{5}$ bands decrease by factors of $\sim$4.5 and $\sim$3.5, respectively (see Fig.~\ref{fig_temp}(b)).
The spectral range considered in this experiment did not include the lowest-energy external mode $\omega_1$.

The large reduction in linewidth of the $\omega_4$ and $\omega_5$ bands that accompanies CDW formation indicates a strong coupling between these vibrations and the electronic states associated with the CDW.
As mentioned, CDW modulation in ZrTe$_3$ opens a gap $2\Delta$ in the dispersion of the electronic band arising from the $5p$ orbitals of Te(2) and Te(3) ions~\cite{Sto1998,Hoe2009}.
The size of $2\Delta$ was found\cite{Hoe2009,Perucchi2005} to be at least 400~cm$^{-1}$, i.e., larger in energy than the vibrations considered here.
Therefore, the reduction in linewidth of modes $\omega_4$ and $\omega_5$ as the gap is opened reflects a loss of electronic relaxational channels for these particular vibrations.
Such linewidth changes have also been observed for strongly coupled phonons in other correlation gap materials, such as the putative Kondo insulator FeSi and the A-15 superconductor Nb$_{3}$Sn.\cite{Nyh1995,Axe1973}
On the other hand, the lower-energy external vibrations are evidently not strongly coupled to the relevant electronic degrees of freedom, as their linewidths are insensitive to the formation of the CDW (see Fig.~\ref{fig_temp}(b)).
Remarkably, the linewidth of the diatomic vibration $\omega_6$ is also largely insensitive to the electronic gap formation, despite the fact that this mode represents a longitudinal vibration of the dimerized Te(2)-Te(3) chain.

\subsection{Pressure dependence of phonon modes in ZrTe$_{3}$}
\label{sec_pressure}

\begin{figure}[h]
\includegraphics{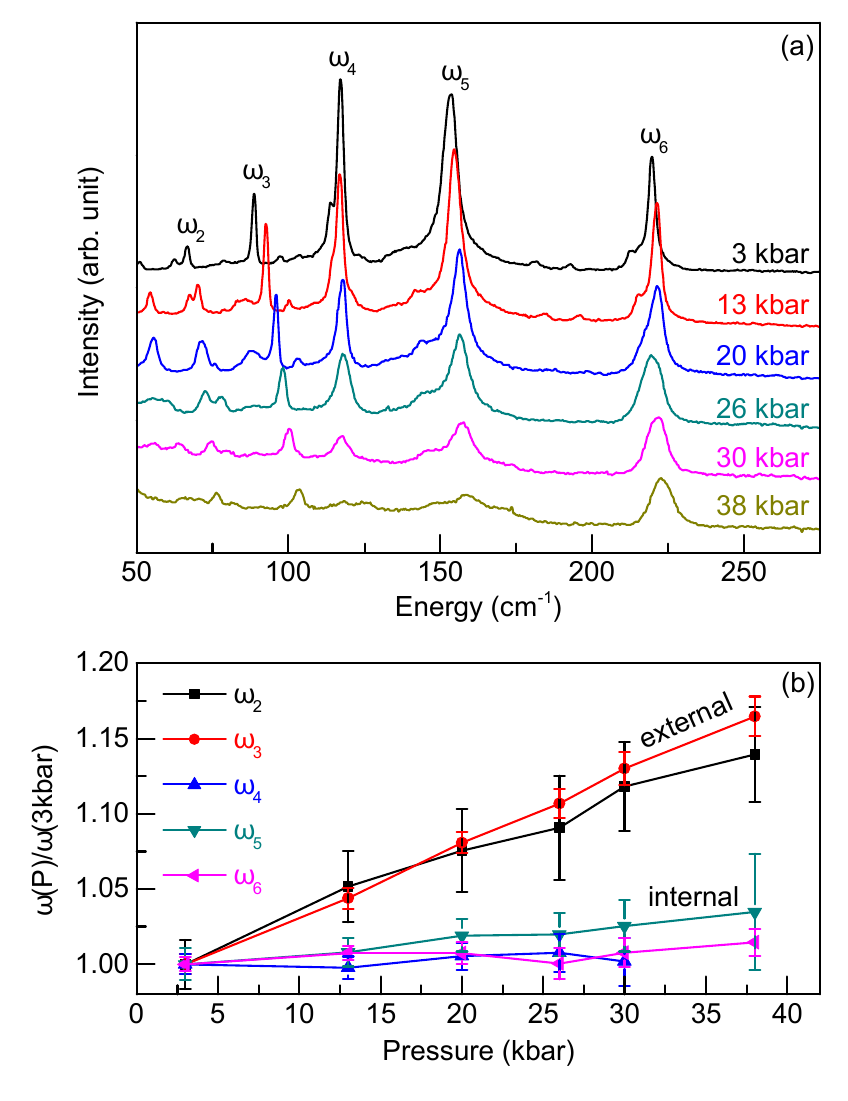}
\caption{\label{fig_pressure} (Color online) (a) Pressure dependence of the ZrTe$_{3}$ Raman spectrum at $T=3$~K.
Spectra have been offset for clarity.
(b) Phonon energies, relative to $P=3$~kbar, as a function of pressure.}
\end{figure}

The pressure dependence of the Raman spectrum of ZrTe$_{3}$ is shown for $T = 3$~K in Fig.~\ref{fig_pressure}(a), with the pressure dependences of the $\omega_2$--$\omega_6$ mode frequencies summarized in Fig.~\ref{fig_pressure}(b).
The frequencies of internal modes $\omega_4$--$\omega_6$ show relatively little sensitivity to applied pressure, while the frequencies of external modes $\omega_2$ and $\omega_3$, which stretch the interchain Zr-Te bonds, increase linearly as a function of pressure.
This indicates that interchain coupling becomes stronger at high pressures, or equivalently, that the crystal becomes more three-dimensional.
Increased three-dimensionality and the resultant reduction in nested Fermi surface has been speculated to contribute to the pressure-induced suppression of the CDW above $\sim$20~kbar in ZrTe$_3$.\cite{PhysRevB.71.132508}
Similar pressure-induced transitions between CDW and superconducting phases caused by enhanced interchain coupling have also been observed in the quasi-1D chain materials NbSe$_{3}$ and Nb$_{3}$Te$_{4}$.\cite{Ido1990,Tan2000}
The decreased one-dimensionality of ZrTe$_{3}$ with pressure also suggests that pressure-induced superconductivity in ZrTe$_{3}$ will not have the filamentary nature that characterizes the ambient pressure superconductivity.\cite{Nak1986,Yam2012}

\begin{figure}[h]
\includegraphics{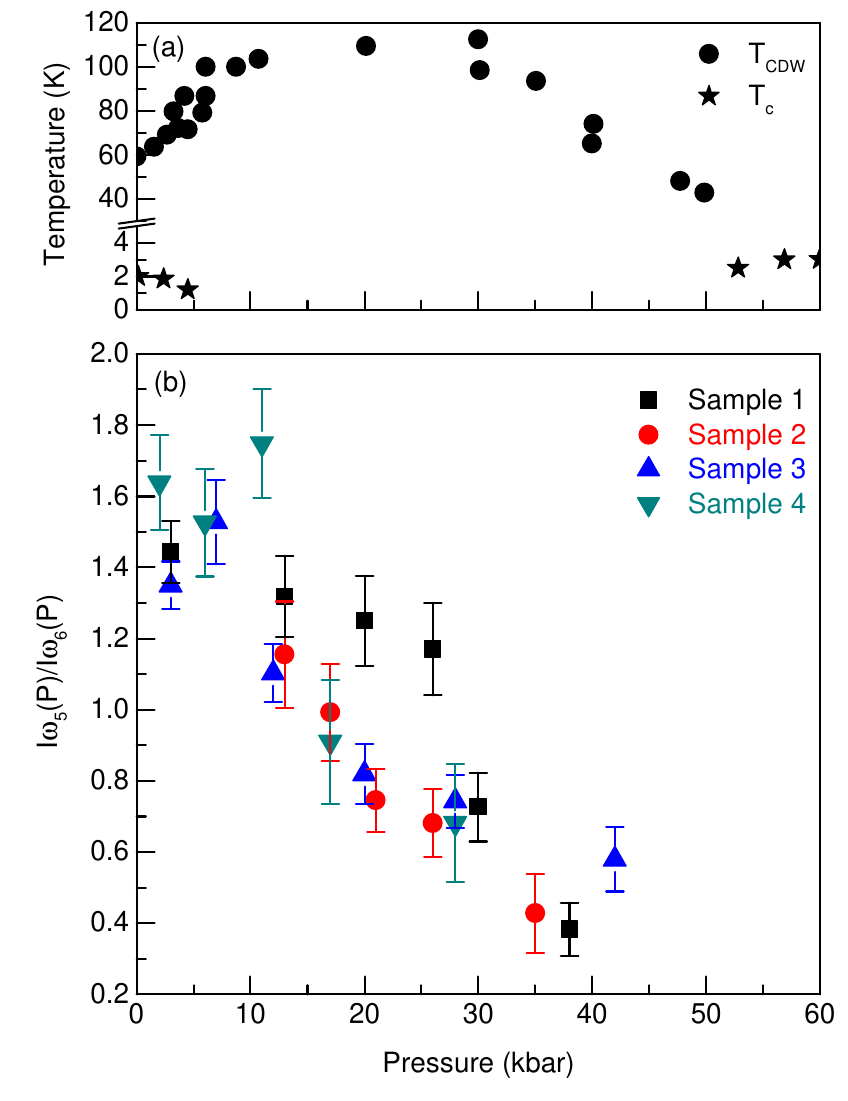}
\caption{\label{fig_intensity} (Color online) (a) Pressure dependence of the CDW transition temperature $T_{CDW}$ and the superconducting transition temperature $T_c$, from Ref.~\onlinecite{PhysRevB.71.132508}.
(b) Pressure dependence of the peak intensity of intra-prismatic-chain mode $\omega_5$ relative to mode $\omega_6$.
Data shown are from measurements of four different samples.}
\end{figure}

The most striking feature of the spectra shown in Fig.~\ref{fig_pressure}(a) is the suppression of internal vibration peaks $\omega_4$ and $\omega_5$ with applied pressure.
Whereas warming through the CDW transition broadens peaks $\omega_4$ and $\omega_5$ by a factor of 3.5--4.5 due to an increase in carriers, increasing pressure has a much smaller ($<$50\%) effect on the linewidths of these peaks.
The pressure-induced loss of integrated intensity of peaks $\omega_4$ and $\omega_5$ is therefore most likely due to a loss of long-range structural order, rather than overdamping due to electron-phonon coupling.
To be more specific, we note that the suppressed modes $\omega_4$ and $\omega_5$ primarily stretch intraprism Zr-Te bonds.
On the other hand, the lower-energy external vibrations $\omega_2$ and $\omega_3$, which primarily involve interprism Zr-Te bonds, as well as the diatomic vibration $\omega_6$ of the Te(2)-Te(3) bond, all persist up to the highest pressures measured.
Therefore, the pressure-induced loss of long-range order and suppression of modes $\omega_4$ and $\omega_5$ is likely due to disorder amongst the intraprism Zr-Te bonds.

Figure~\ref{fig_intensity}(b) shows the pressure-dependence of the peak intensity of the $\omega_5$ band, relative to the robust $\omega_6$ band.
For comparison, Fig.~\ref{fig_intensity}(a) shows the evolution of $T_{CDW}$ (filled circles) as a function of pressure.
The peak intensity decreases rapidly above $\sim$10~kbar, and is almost completely suppressed at pressures approaching the CDW collapse at $\sim$50~kbar.
Since the $\omega_4$ and $\omega_5$ vibrations have already been shown to be strongly coupled to the electronic degrees of freedom involved in CDW formation, and since the suppression of $T_{CDW}$ follows the suppression of modes $\omega_4$ and $\omega_5$, it is natural to associate the disorder amongst intraprism Zr-Te bonds with the collapse of the CDW.
In support of this interpretation, we note that a recent \emph{ambient}-pressure study of single-crystal ZrTe$_3$ grown at elevated temperatures also observed suppression of the CDW in favor of superconductivity.\cite{Zhu2013}
This was attributed to random displacements of the Zr and Te(1) atoms along the $c$ axis as measured with x-ray diffraction, which is an interpretation quite similar to the pressure-induced structural changes proposed in our study.
High-pressure x-ray-diffraction measurements of pure ZrTe$_3$ would be useful for confirming this interpretation of our high-pressure results.

The pressure dependences of the ZrTe$_3$ optical phonons demonstrate not only a trend toward higher-dimensionality, which may explain the pressure-induced suppression of the CDW in general terms via a loss of nested Fermi surface, but also indicate a specific structural change that appears to precede CDW collapse; namely, a loss of long-range order in the ZrTe$_3$ trigonal prismatic chains.

\section{Summary}

We have performed variable-temperature Raman measurements of ZrTe$_3$ which show that certain ZrTe$_3$-chain-deforming vibrations exhibit dramatic linewidth reductions that accompany CDW formation in this crystal, indicating that these vibrations are strongly coupled to the electronic states involved in the formation of the CDW.
We have also performed variable-pressure Raman measurements which show that the same ZrTe$_3$-chain-deforming vibrations are strongly suppressed at pressures above $\sim$10~kbar, mirroring the previously-observed pressure-induced suppression of the CDW.
We argue that this suppression reflects a loss of long-range structural order at high pressures, specifically amongst the intrachain Zr-Te bonds, and that such a structural degredation may be responsible for the eventual pressure-induced collapse of the CDW.

\begin{acknowledgments}
Research was supported by the U.S. Department of Energy, Office of Basic Energy Sciences, Division of Materials Sciences and Engineering under Award DE-FG02-07ER46453 and by the National Science Foundation under Grant NSF DMR 08-56321 (Y.~Gim).
T.~Byrum was partially supported by the National Science Foundation Graduate Research Fellowship Program under Grant No. DGE-1144245.
Work at Brookhaven National Laboratory was supported by the U.S. Department of Energy under contract No.~DE-AC02-98CH10886.
\end{acknowledgments}


\begin{thebibliography}{33}%
\makeatletter
\providecommand \@ifxundefined [1]{%
 \@ifx{#1\undefined}
}%
\providecommand \@ifnum [1]{%
 \ifnum #1\expandafter \@firstoftwo
 \else \expandafter \@secondoftwo
 \fi
}%
\providecommand \@ifx [1]{%
 \ifx #1\expandafter \@firstoftwo
 \else \expandafter \@secondoftwo
 \fi
}%
\providecommand \natexlab [1]{#1}%
\providecommand \enquote  [1]{``#1''}%
\providecommand \bibnamefont  [1]{#1}%
\providecommand \bibfnamefont [1]{#1}%
\providecommand \citenamefont [1]{#1}%
\providecommand \href@noop [0]{\@secondoftwo}%
\providecommand \href [0]{\begingroup \@sanitize@url \@href}%
\providecommand \@href[1]{\@@startlink{#1}\@@href}%
\providecommand \@@href[1]{\endgroup#1\@@endlink}%
\providecommand \@sanitize@url [0]{\catcode `\\12\catcode `\$12\catcode
  `\&12\catcode `\#12\catcode `\^12\catcode `\_12\catcode `\%12\relax}%
\providecommand \@@startlink[1]{}%
\providecommand \@@endlink[0]{}%
\providecommand \url  [0]{\begingroup\@sanitize@url \@url }%
\providecommand \@url [1]{\endgroup\@href {#1}{\urlprefix }}%
\providecommand \urlprefix  [0]{URL }%
\providecommand \Eprint [0]{\href }%
\providecommand \doibase [0]{http://dx.doi.org/}%
\providecommand \selectlanguage [0]{\@gobble}%
\providecommand \bibinfo  [0]{\@secondoftwo}%
\providecommand \bibfield  [0]{\@secondoftwo}%
\providecommand \translation [1]{[#1]}%
\providecommand \BibitemOpen [0]{}%
\providecommand \bibitemStop [0]{}%
\providecommand \bibitemNoStop [0]{.\EOS\space}%
\providecommand \EOS [0]{\spacefactor3000\relax}%
\providecommand \BibitemShut  [1]{\csname bibitem#1\endcsname}%
\let\auto@bib@innerbib\@empty
\bibitem [{\citenamefont {Kivelson}\ and\ \citenamefont
  {Fradkin}(2007)}]{Kiv.2007}%
  \BibitemOpen
  \bibfield  {author} {\bibinfo {author} {\bibfnamefont {S.}~\bibnamefont
  {Kivelson}}\ and\ \bibinfo {author} {\bibfnamefont {E.}~\bibnamefont
  {Fradkin}},\ }in\ \href {\doibase 10.1007/978-0-387-68734-6_15} {\emph
  {\bibinfo {booktitle} {Handbook of High-Temperature Superconductivity}}},\
  \bibinfo {editor} {edited by\ \bibinfo {editor} {\bibfnamefont
  {J.}~\bibnamefont {Schrieffer}}\ and\ \bibinfo {editor} {\bibfnamefont
  {J.}~\bibnamefont {Brooks}}}\ (\bibinfo  {publisher} {Springer New York},\
  \bibinfo {year} {2007})\ pp.\ \bibinfo {pages} {570--596}\BibitemShut
  {NoStop}%
\bibitem [{\citenamefont {Canfield}\ and\ \citenamefont
  {Bud'ko}(2010)}]{doi:10.1146/annurev-conmatphys-070909-104041}%
  \BibitemOpen
  \bibfield  {author} {\bibinfo {author} {\bibfnamefont {P.~C.}\ \bibnamefont
  {Canfield}}\ and\ \bibinfo {author} {\bibfnamefont {S.~L.}\ \bibnamefont
  {Bud'ko}},\ }\href {\doibase 10.1146/annurev-conmatphys-070909-104041}
  {\bibfield  {journal} {\bibinfo  {journal} {Annual Review of Condensed Matter
  Physics}\ }\textbf {\bibinfo {volume} {1}},\ \bibinfo {pages} {27} (\bibinfo
  {year} {2010})},\ \Eprint
  {http://arxiv.org/abs/http://dx.doi.org/10.1146/annurev-conmatphys-070909-104041}
  {http://dx.doi.org/10.1146/annurev-conmatphys-070909-104041} \BibitemShut
  {NoStop}%
\bibitem [{\citenamefont {Kusmartseva}\ \emph {et~al.}(2009)\citenamefont
  {Kusmartseva}, \citenamefont {Sipos}, \citenamefont {Berger}, \citenamefont
  {Forr\'o},\ and\ \citenamefont {Tuti\ifmmode~\check{s}\else
  \v{s}\fi{}}}]{PhysRevLett.103.236401}%
  \BibitemOpen
  \bibfield  {author} {\bibinfo {author} {\bibfnamefont {A.~F.}\ \bibnamefont
  {Kusmartseva}}, \bibinfo {author} {\bibfnamefont {B.}~\bibnamefont {Sipos}},
  \bibinfo {author} {\bibfnamefont {H.}~\bibnamefont {Berger}}, \bibinfo
  {author} {\bibfnamefont {L.}~\bibnamefont {Forr\'o}}, \ and\ \bibinfo
  {author} {\bibfnamefont {E.}~\bibnamefont {Tuti\ifmmode~\check{s}\else
  \v{s}\fi{}}},\ }\href {\doibase 10.1103/PhysRevLett.103.236401} {\bibfield
  {journal} {\bibinfo  {journal} {Phys. Rev. Lett.}\ }\textbf {\bibinfo
  {volume} {103}},\ \bibinfo {pages} {236401} (\bibinfo {year}
  {2009})}\BibitemShut {NoStop}%
\bibitem [{\citenamefont {Morosan}\ \emph {et~al.}(2006)\citenamefont
  {Morosan}, \citenamefont {Zandbergen}, \citenamefont {Dennis}, \citenamefont
  {Bos}, \citenamefont {Onose}, \citenamefont {Klimczuk}, \citenamefont
  {Ramirez}, \citenamefont {Ong},\ and\ \citenamefont {Cava}}]{Mor2006}%
  \BibitemOpen
  \bibfield  {author} {\bibinfo {author} {\bibfnamefont {E.}~\bibnamefont
  {Morosan}}, \bibinfo {author} {\bibfnamefont {H.~W.}\ \bibnamefont
  {Zandbergen}}, \bibinfo {author} {\bibfnamefont {B.~S.}\ \bibnamefont
  {Dennis}}, \bibinfo {author} {\bibfnamefont {J.~W.~G.}\ \bibnamefont {Bos}},
  \bibinfo {author} {\bibfnamefont {Y.}~\bibnamefont {Onose}}, \bibinfo
  {author} {\bibfnamefont {T.}~\bibnamefont {Klimczuk}}, \bibinfo {author}
  {\bibfnamefont {A.~P.}\ \bibnamefont {Ramirez}}, \bibinfo {author}
  {\bibfnamefont {N.~P.}\ \bibnamefont {Ong}}, \ and\ \bibinfo {author}
  {\bibfnamefont {R.~J.}\ \bibnamefont {Cava}},\ }\href {\doibase
  10.1038/nphys360} {\bibfield  {journal} {\bibinfo  {journal} {Nature
  Physics}\ }\textbf {\bibinfo {volume} {2}},\ \bibinfo {pages} {544} (\bibinfo
  {year} {2006})}\BibitemShut {NoStop}%
\bibitem [{\citenamefont {Barath}\ \emph {et~al.}(2008)\citenamefont {Barath},
  \citenamefont {Kim}, \citenamefont {Karpus}, \citenamefont {Cooper},
  \citenamefont {Abbamonte}, \citenamefont {Fradkin}, \citenamefont {Morosan},\
  and\ \citenamefont {Cava}}]{Bar2008}%
  \BibitemOpen
  \bibfield  {author} {\bibinfo {author} {\bibfnamefont {H.}~\bibnamefont
  {Barath}}, \bibinfo {author} {\bibfnamefont {M.}~\bibnamefont {Kim}},
  \bibinfo {author} {\bibfnamefont {J.~F.}\ \bibnamefont {Karpus}}, \bibinfo
  {author} {\bibfnamefont {S.~L.}\ \bibnamefont {Cooper}}, \bibinfo {author}
  {\bibfnamefont {P.}~\bibnamefont {Abbamonte}}, \bibinfo {author}
  {\bibfnamefont {E.}~\bibnamefont {Fradkin}}, \bibinfo {author} {\bibfnamefont
  {E.}~\bibnamefont {Morosan}}, \ and\ \bibinfo {author} {\bibfnamefont
  {R.~J.}\ \bibnamefont {Cava}},\ }\href {\doibase
  10.1103/PhysRevLett.100.106402} {\bibfield  {journal} {\bibinfo  {journal}
  {Phys. Rev. Lett.}\ }\textbf {\bibinfo {volume} {100}},\ \bibinfo {pages}
  {106402} (\bibinfo {year} {2008})}\BibitemShut {NoStop}%
\bibitem [{\citenamefont {Joe}\ \emph {et~al.}(2014)\citenamefont {Joe},
  \citenamefont {Chen}, \citenamefont {Ghaemi}, \citenamefont {Finkelstein},
  \citenamefont {de~la Pe\~{n}a}, \citenamefont {Gan}, \citenamefont {Lee},
  \citenamefont {Yuan}, \citenamefont {Geck}, \citenamefont {MacDougall},
  \citenamefont {Chiang}, \citenamefont {Cooper}, \citenamefont {Fradkin},\
  and\ \citenamefont {Abbamonte}}]{Joe2014}%
  \BibitemOpen
  \bibfield  {author} {\bibinfo {author} {\bibfnamefont {Y.~I.}\ \bibnamefont
  {Joe}}, \bibinfo {author} {\bibfnamefont {X.~M.}\ \bibnamefont {Chen}},
  \bibinfo {author} {\bibfnamefont {P.}~\bibnamefont {Ghaemi}}, \bibinfo
  {author} {\bibfnamefont {K.~D.}\ \bibnamefont {Finkelstein}}, \bibinfo
  {author} {\bibfnamefont {G.~a.}\ \bibnamefont {de~la Pe\~{n}a}}, \bibinfo
  {author} {\bibfnamefont {Y.}~\bibnamefont {Gan}}, \bibinfo {author}
  {\bibfnamefont {J.~C.~T.}\ \bibnamefont {Lee}}, \bibinfo {author}
  {\bibfnamefont {S.}~\bibnamefont {Yuan}}, \bibinfo {author} {\bibfnamefont
  {J.}~\bibnamefont {Geck}}, \bibinfo {author} {\bibfnamefont {G.~J.}\
  \bibnamefont {MacDougall}}, \bibinfo {author} {\bibfnamefont {T.~C.}\
  \bibnamefont {Chiang}}, \bibinfo {author} {\bibfnamefont {S.~L.}\
  \bibnamefont {Cooper}}, \bibinfo {author} {\bibfnamefont {E.}~\bibnamefont
  {Fradkin}}, \ and\ \bibinfo {author} {\bibfnamefont {P.}~\bibnamefont
  {Abbamonte}},\ }\href {\doibase 10.1038/nphys2935} {\bibfield  {journal}
  {\bibinfo  {journal} {Nature Physics}\ }\textbf {\bibinfo {volume} {10}},\
  \bibinfo {pages} {421} (\bibinfo {year} {2014})}\BibitemShut {NoStop}%
\bibitem [{\citenamefont {Wagner}\ \emph {et~al.}(2008)\citenamefont {Wagner},
  \citenamefont {Morosan}, \citenamefont {Hor}, \citenamefont {Tao},
  \citenamefont {Zhu}, \citenamefont {Sanders}, \citenamefont {McQueen},
  \citenamefont {Zandbergen}, \citenamefont {Williams}, \citenamefont {West},\
  and\ \citenamefont {Cava}}]{PhysRevB.78.104520}%
  \BibitemOpen
  \bibfield  {author} {\bibinfo {author} {\bibfnamefont {K.~E.}\ \bibnamefont
  {Wagner}}, \bibinfo {author} {\bibfnamefont {E.}~\bibnamefont {Morosan}},
  \bibinfo {author} {\bibfnamefont {Y.~S.}\ \bibnamefont {Hor}}, \bibinfo
  {author} {\bibfnamefont {J.}~\bibnamefont {Tao}}, \bibinfo {author}
  {\bibfnamefont {Y.}~\bibnamefont {Zhu}}, \bibinfo {author} {\bibfnamefont
  {T.}~\bibnamefont {Sanders}}, \bibinfo {author} {\bibfnamefont {T.~M.}\
  \bibnamefont {McQueen}}, \bibinfo {author} {\bibfnamefont {H.~W.}\
  \bibnamefont {Zandbergen}}, \bibinfo {author} {\bibfnamefont {A.~J.}\
  \bibnamefont {Williams}}, \bibinfo {author} {\bibfnamefont {D.~V.}\
  \bibnamefont {West}}, \ and\ \bibinfo {author} {\bibfnamefont {R.~J.}\
  \bibnamefont {Cava}},\ }\href {\doibase 10.1103/PhysRevB.78.104520}
  {\bibfield  {journal} {\bibinfo  {journal} {Phys. Rev. B}\ }\textbf {\bibinfo
  {volume} {78}},\ \bibinfo {pages} {104520} (\bibinfo {year}
  {2008})}\BibitemShut {NoStop}%
\bibitem [{\citenamefont {Lei}\ \emph {et~al.}(2011)\citenamefont {Lei},
  \citenamefont {Zhu},\ and\ \citenamefont {Petrovic}}]{Lei2011}%
  \BibitemOpen
  \bibfield  {author} {\bibinfo {author} {\bibfnamefont {H.}~\bibnamefont
  {Lei}}, \bibinfo {author} {\bibfnamefont {X.}~\bibnamefont {Zhu}}, \ and\
  \bibinfo {author} {\bibfnamefont {C.}~\bibnamefont {Petrovic}},\ }\href@noop
  {} {\bibfield  {journal} {\bibinfo  {journal} {EPL (Europhysics Letters)}\
  }\textbf {\bibinfo {volume} {95}},\ \bibinfo {pages} {17011} (\bibinfo {year}
  {2011})}\BibitemShut {NoStop}%
\bibitem [{\citenamefont {Zhu}\ \emph {et~al.}(2011)\citenamefont {Zhu},
  \citenamefont {Lei},\ and\ \citenamefont
  {Petrovic}}]{PhysRevLett.106.246404}%
  \BibitemOpen
  \bibfield  {author} {\bibinfo {author} {\bibfnamefont {X.}~\bibnamefont
  {Zhu}}, \bibinfo {author} {\bibfnamefont {H.}~\bibnamefont {Lei}}, \ and\
  \bibinfo {author} {\bibfnamefont {C.}~\bibnamefont {Petrovic}},\ }\href
  {\doibase 10.1103/PhysRevLett.106.246404} {\bibfield  {journal} {\bibinfo
  {journal} {Phys. Rev. Lett.}\ }\textbf {\bibinfo {volume} {106}},\ \bibinfo
  {pages} {246404} (\bibinfo {year} {2011})}\BibitemShut {NoStop}%
\bibitem [{\citenamefont {Yomo}\ \emph {et~al.}(2005)\citenamefont {Yomo},
  \citenamefont {Yamaya}, \citenamefont {Abliz}, \citenamefont {Hedo},\ and\
  \citenamefont {Uwatoko}}]{PhysRevB.71.132508}%
  \BibitemOpen
  \bibfield  {author} {\bibinfo {author} {\bibfnamefont {R.}~\bibnamefont
  {Yomo}}, \bibinfo {author} {\bibfnamefont {K.}~\bibnamefont {Yamaya}},
  \bibinfo {author} {\bibfnamefont {M.}~\bibnamefont {Abliz}}, \bibinfo
  {author} {\bibfnamefont {M.}~\bibnamefont {Hedo}}, \ and\ \bibinfo {author}
  {\bibfnamefont {Y.}~\bibnamefont {Uwatoko}},\ }\href {\doibase
  10.1103/PhysRevB.71.132508} {\bibfield  {journal} {\bibinfo  {journal} {Phys.
  Rev. B}\ }\textbf {\bibinfo {volume} {71}},\ \bibinfo {pages} {132508}
  (\bibinfo {year} {2005})}\BibitemShut {NoStop}%
\bibitem [{\citenamefont {St\H{o}we}\ and\ \citenamefont
  {Wagner}(1998)}]{Sto1998}%
  \BibitemOpen
  \bibfield  {author} {\bibinfo {author} {\bibfnamefont {K.}~\bibnamefont
  {St\H{o}we}}\ and\ \bibinfo {author} {\bibfnamefont {F.~R.}\ \bibnamefont
  {Wagner}},\ }\href {\doibase http://dx.doi.org/10.1006/jssc.1998.7769}
  {\bibfield  {journal} {\bibinfo  {journal} {Journal of Solid State
  Chemistry}\ }\textbf {\bibinfo {volume} {138}},\ \bibinfo {pages} {160 }
  (\bibinfo {year} {1998})}\BibitemShut {NoStop}%
\bibitem [{\citenamefont {Takahashi}\ \emph {et~al.}(1984)\citenamefont
  {Takahashi}, \citenamefont {Sambongi}, \citenamefont {Brill},\ and\
  \citenamefont {Roark}}]{Tak1984}%
  \BibitemOpen
  \bibfield  {author} {\bibinfo {author} {\bibfnamefont {S.}~\bibnamefont
  {Takahashi}}, \bibinfo {author} {\bibfnamefont {T.}~\bibnamefont {Sambongi}},
  \bibinfo {author} {\bibfnamefont {J.}~\bibnamefont {Brill}}, \ and\ \bibinfo
  {author} {\bibfnamefont {W.}~\bibnamefont {Roark}},\ }\href {\doibase
  http://dx.doi.org/10.1016/0038-1098(84)90416-2} {\bibfield  {journal}
  {\bibinfo  {journal} {Solid State Communications}\ }\textbf {\bibinfo
  {volume} {49}},\ \bibinfo {pages} {1031 } (\bibinfo {year}
  {1984})}\BibitemShut {NoStop}%
\bibitem [{\citenamefont {Nakajima}\ \emph {et~al.}(1986)\citenamefont
  {Nakajima}, \citenamefont {Nomura},\ and\ \citenamefont
  {Sambongi}}]{Nak1986}%
  \BibitemOpen
  \bibfield  {author} {\bibinfo {author} {\bibfnamefont {H.}~\bibnamefont
  {Nakajima}}, \bibinfo {author} {\bibfnamefont {K.}~\bibnamefont {Nomura}}, \
  and\ \bibinfo {author} {\bibfnamefont {T.}~\bibnamefont {Sambongi}},\ }\href
  {\doibase http://dx.doi.org/10.1016/0378-4363(86)90106-3} {\bibfield
  {journal} {\bibinfo  {journal} {Physica B+C}\ }\textbf {\bibinfo {volume}
  {143}},\ \bibinfo {pages} {240 } (\bibinfo {year} {1986})}\BibitemShut
  {NoStop}%
\bibitem [{\citenamefont {{Eaglesham}}\ \emph {et~al.}(1984)\citenamefont
  {{Eaglesham}}, \citenamefont {{Steeds}},\ and\ \citenamefont
  {{Wilson}}}]{Eag1984}%
  \BibitemOpen
  \bibfield  {author} {\bibinfo {author} {\bibfnamefont {D.~J.}\ \bibnamefont
  {{Eaglesham}}}, \bibinfo {author} {\bibfnamefont {J.~W.}\ \bibnamefont
  {{Steeds}}}, \ and\ \bibinfo {author} {\bibfnamefont {J.~A.}\ \bibnamefont
  {{Wilson}}},\ }\href {\doibase 10.1088/0022-3719/17/27/001} {\bibfield
  {journal} {\bibinfo  {journal} {Journal of Physics C Solid State Physics}\
  }\textbf {\bibinfo {volume} {17}},\ \bibinfo {pages} {L697} (\bibinfo {year}
  {1984})}\BibitemShut {NoStop}%
\bibitem [{\citenamefont {Hoesch}\ \emph
  {et~al.}(2009{\natexlab{a}})\citenamefont {Hoesch}, \citenamefont {Bosak},
  \citenamefont {Chernyshov}, \citenamefont {Berger},\ and\ \citenamefont
  {Krisch}}]{Hoe2009.2}%
  \BibitemOpen
  \bibfield  {author} {\bibinfo {author} {\bibfnamefont {M.}~\bibnamefont
  {Hoesch}}, \bibinfo {author} {\bibfnamefont {A.}~\bibnamefont {Bosak}},
  \bibinfo {author} {\bibfnamefont {D.}~\bibnamefont {Chernyshov}}, \bibinfo
  {author} {\bibfnamefont {H.}~\bibnamefont {Berger}}, \ and\ \bibinfo {author}
  {\bibfnamefont {M.}~\bibnamefont {Krisch}},\ }\href {\doibase
  10.1103/PhysRevLett.102.086402} {\bibfield  {journal} {\bibinfo  {journal}
  {Phys. Rev. Lett.}\ }\textbf {\bibinfo {volume} {102}},\ \bibinfo {pages}
  {086402} (\bibinfo {year} {2009}{\natexlab{a}})}\BibitemShut {NoStop}%
\bibitem [{\citenamefont {Yokoya}\ \emph {et~al.}(2005)\citenamefont {Yokoya},
  \citenamefont {Kiss}, \citenamefont {Chainani}, \citenamefont {Shin},\ and\
  \citenamefont {Yamaya}}]{Yok2005}%
  \BibitemOpen
  \bibfield  {author} {\bibinfo {author} {\bibfnamefont {T.}~\bibnamefont
  {Yokoya}}, \bibinfo {author} {\bibfnamefont {T.}~\bibnamefont {Kiss}},
  \bibinfo {author} {\bibfnamefont {A.}~\bibnamefont {Chainani}}, \bibinfo
  {author} {\bibfnamefont {S.}~\bibnamefont {Shin}}, \ and\ \bibinfo {author}
  {\bibfnamefont {K.}~\bibnamefont {Yamaya}},\ }\href {\doibase
  10.1103/PhysRevB.71.140504} {\bibfield  {journal} {\bibinfo  {journal} {Phys.
  Rev. B}\ }\textbf {\bibinfo {volume} {71}},\ \bibinfo {pages} {140504}
  (\bibinfo {year} {2005})}\BibitemShut {NoStop}%
\bibitem [{\citenamefont {Perucchi}\ \emph {et~al.}(2005)\citenamefont
  {Perucchi}, \citenamefont {Degiorgi},\ and\ \citenamefont
  {Berger}}]{Perucchi2005}%
  \BibitemOpen
  \bibfield  {author} {\bibinfo {author} {\bibfnamefont {A.}~\bibnamefont
  {Perucchi}}, \bibinfo {author} {\bibfnamefont {L.}~\bibnamefont {Degiorgi}},
  \ and\ \bibinfo {author} {\bibfnamefont {H.}~\bibnamefont {Berger}},\ }\href
  {http://search.ebscohost.com/login.aspx?direct=true&db=a9h&AN=19439399&site=ehost-live}
  {\bibfield  {journal} {\bibinfo  {journal} {European Physical Journal B --
  Condensed Matter}\ }\textbf {\bibinfo {volume} {48}},\ \bibinfo {pages} {489
  } (\bibinfo {year} {2005})}\BibitemShut {NoStop}%
\bibitem [{\citenamefont {Felser}\ \emph {et~al.}(1998)\citenamefont {Felser},
  \citenamefont {W.~Finckh}, \citenamefont {Kleinke}, \citenamefont {Rocker},\
  and\ \citenamefont {Tremel}}]{Fel1998}%
  \BibitemOpen
  \bibfield  {author} {\bibinfo {author} {\bibfnamefont {C.}~\bibnamefont
  {Felser}}, \bibinfo {author} {\bibfnamefont {E.}~\bibnamefont {W.~Finckh}},
  \bibinfo {author} {\bibfnamefont {H.}~\bibnamefont {Kleinke}}, \bibinfo
  {author} {\bibfnamefont {F.}~\bibnamefont {Rocker}}, \ and\ \bibinfo {author}
  {\bibfnamefont {W.}~\bibnamefont {Tremel}},\ }\href {\doibase
  10.1039/A802948B} {\bibfield  {journal} {\bibinfo  {journal} {J. Mater.
  Chem.}\ }\textbf {\bibinfo {volume} {8}},\ \bibinfo {pages} {1787} (\bibinfo
  {year} {1998})}\BibitemShut {NoStop}%
\bibitem [{\citenamefont {Snow}\ \emph {et~al.}(2003)\citenamefont {Snow},
  \citenamefont {Karpus}, \citenamefont {Cooper}, \citenamefont {Kidd},\ and\
  \citenamefont {Chiang}}]{Sno2003}%
  \BibitemOpen
  \bibfield  {author} {\bibinfo {author} {\bibfnamefont {C.~S.}\ \bibnamefont
  {Snow}}, \bibinfo {author} {\bibfnamefont {J.~F.}\ \bibnamefont {Karpus}},
  \bibinfo {author} {\bibfnamefont {S.~L.}\ \bibnamefont {Cooper}}, \bibinfo
  {author} {\bibfnamefont {T.~E.}\ \bibnamefont {Kidd}}, \ and\ \bibinfo
  {author} {\bibfnamefont {T.-C.}\ \bibnamefont {Chiang}},\ }\href {\doibase
  10.1103/PhysRevLett.91.136402} {\bibfield  {journal} {\bibinfo  {journal}
  {Phys. Rev. Lett.}\ }\textbf {\bibinfo {volume} {91}},\ \bibinfo {pages}
  {136402} (\bibinfo {year} {2003})}\BibitemShut {NoStop}%
\bibitem [{\citenamefont {Block}\ and\ \citenamefont
  {Piermarini}(1976)}]{Block1976}%
  \BibitemOpen
  \bibfield  {author} {\bibinfo {author} {\bibfnamefont {S.}~\bibnamefont
  {Block}}\ and\ \bibinfo {author} {\bibfnamefont {G.}~\bibnamefont
  {Piermarini}},\ }\href
  {http://search.ebscohost.com/login.aspx?direct=true&db=a9h&AN=4958492&site=ehost-live}
  {\bibfield  {journal} {\bibinfo  {journal} {Physics Today}\ }\textbf
  {\bibinfo {volume} {29}},\ \bibinfo {pages} {44} (\bibinfo {year}
  {1976})}\BibitemShut {NoStop}%
\bibitem [{\citenamefont {Zwick}\ and\ \citenamefont
  {Renucci}(1979)}]{Zwi1979}%
  \BibitemOpen
  \bibfield  {author} {\bibinfo {author} {\bibfnamefont {A.}~\bibnamefont
  {Zwick}}\ and\ \bibinfo {author} {\bibfnamefont {M.~A.}\ \bibnamefont
  {Renucci}},\ }\href {\doibase 10.1002/pssb.2220960232} {\bibfield  {journal}
  {\bibinfo  {journal} {physica status solidi (b)}\ }\textbf {\bibinfo {volume}
  {96}},\ \bibinfo {pages} {757} (\bibinfo {year} {1979})}\BibitemShut
  {NoStop}%
\bibitem [{\citenamefont {Grisel}\ \emph {et~al.}(1980)\citenamefont {Grisel},
  \citenamefont {L\'{e}vy},\ and\ \citenamefont {Wieting}}]{Wieting1980}%
  \BibitemOpen
  \bibfield  {author} {\bibinfo {author} {\bibfnamefont {A.}~\bibnamefont
  {Grisel}}, \bibinfo {author} {\bibfnamefont {F.}~\bibnamefont {L\'{e}vy}}, \
  and\ \bibinfo {author} {\bibfnamefont {T.}~\bibnamefont {Wieting}},\ }\href
  {\doibase http://dx.doi.org/10.1016/0378-4363(80)90262-4} {\bibfield
  {journal} {\bibinfo  {journal} {Physica B+C}\ }\textbf {\bibinfo {volume}
  {99}},\ \bibinfo {pages} {365 } (\bibinfo {year} {1980})}\BibitemShut
  {NoStop}%
\bibitem [{\citenamefont {{Zwick}}\ \emph {et~al.}(1980)\citenamefont
  {{Zwick}}, \citenamefont {{Renucci}},\ and\ \citenamefont
  {{Kjekshus}}}]{Zwi1980}%
  \BibitemOpen
  \bibfield  {author} {\bibinfo {author} {\bibfnamefont {A.}~\bibnamefont
  {{Zwick}}}, \bibinfo {author} {\bibfnamefont {M.~A.}\ \bibnamefont
  {{Renucci}}}, \ and\ \bibinfo {author} {\bibfnamefont {A.}~\bibnamefont
  {{Kjekshus}}},\ }\href {\doibase 10.1088/0022-3719/13/30/023} {\bibfield
  {journal} {\bibinfo  {journal} {Journal of Physics C Solid State Physics}\
  }\textbf {\bibinfo {volume} {13}},\ \bibinfo {pages} {5603} (\bibinfo {year}
  {1980})}\BibitemShut {NoStop}%
\bibitem [{\citenamefont {Wieting}\ \emph {et~al.}(1981)\citenamefont
  {Wieting}, \citenamefont {Grisel},\ and\ \citenamefont
  {L\'{e}vy}}]{Wieting1981}%
  \BibitemOpen
  \bibfield  {author} {\bibinfo {author} {\bibfnamefont {T.}~\bibnamefont
  {Wieting}}, \bibinfo {author} {\bibfnamefont {A.}~\bibnamefont {Grisel}}, \
  and\ \bibinfo {author} {\bibfnamefont {F.}~\bibnamefont {L\'{e}vy}},\ }\href
  {\doibase http://dx.doi.org/10.1016/0378-4363(81)90277-1} {\bibfield
  {journal} {\bibinfo  {journal} {Physica B+C}\ }\textbf {\bibinfo {volume}
  {105}},\ \bibinfo {pages} {366 } (\bibinfo {year} {1981})}\BibitemShut
  {NoStop}%
\bibitem [{\citenamefont {Fateley}\ \emph {et~al.}(1972)\citenamefont
  {Fateley}, \citenamefont {Dollish}, \citenamefont {McDevitt},\ and\
  \citenamefont {Bentley}}]{Fateley1972}%
  \BibitemOpen
  \bibfield  {author} {\bibinfo {author} {\bibfnamefont {W.~G.}\ \bibnamefont
  {Fateley}}, \bibinfo {author} {\bibfnamefont {F.~R.}\ \bibnamefont
  {Dollish}}, \bibinfo {author} {\bibfnamefont {N.~T.}\ \bibnamefont
  {McDevitt}}, \ and\ \bibinfo {author} {\bibfnamefont {F.~F.}\ \bibnamefont
  {Bentley}},\ }\href@noop {} {\emph {\bibinfo {title} {Infrared and Raman
  Selection Rules for Molecular and Lattice Vibrations: The Correlation
  Method}}}\ (\bibinfo {year} {1972})\BibitemShut {NoStop}%
\bibitem [{\citenamefont {Fano}(1961)}]{Fan1866}%
  \BibitemOpen
  \bibfield  {author} {\bibinfo {author} {\bibfnamefont {U.}~\bibnamefont
  {Fano}},\ }\href {\doibase 10.1103/PhysRev.124.1866} {\bibfield  {journal}
  {\bibinfo  {journal} {Phys. Rev.}\ }\textbf {\bibinfo {volume} {124}},\
  \bibinfo {pages} {1866} (\bibinfo {year} {1961})}\BibitemShut {NoStop}%
\bibitem [{\citenamefont {Hoesch}\ \emph
  {et~al.}(2009{\natexlab{b}})\citenamefont {Hoesch}, \citenamefont {Cui},
  \citenamefont {Shimada}, \citenamefont {Battaglia}, \citenamefont
  {Fujimori},\ and\ \citenamefont {Berger}}]{Hoe2009}%
  \BibitemOpen
  \bibfield  {author} {\bibinfo {author} {\bibfnamefont {M.}~\bibnamefont
  {Hoesch}}, \bibinfo {author} {\bibfnamefont {X.}~\bibnamefont {Cui}},
  \bibinfo {author} {\bibfnamefont {K.}~\bibnamefont {Shimada}}, \bibinfo
  {author} {\bibfnamefont {C.}~\bibnamefont {Battaglia}}, \bibinfo {author}
  {\bibfnamefont {S.-i.}\ \bibnamefont {Fujimori}}, \ and\ \bibinfo {author}
  {\bibfnamefont {H.}~\bibnamefont {Berger}},\ }\href {\doibase
  10.1103/PhysRevB.80.075423} {\bibfield  {journal} {\bibinfo  {journal} {Phys.
  Rev. B}\ }\textbf {\bibinfo {volume} {80}},\ \bibinfo {pages} {075423}
  (\bibinfo {year} {2009}{\natexlab{b}})}\BibitemShut {NoStop}%
\bibitem [{\citenamefont {Nyhus}\ \emph {et~al.}(1995)\citenamefont {Nyhus},
  \citenamefont {Cooper},\ and\ \citenamefont {Fisk}}]{Nyh1995}%
  \BibitemOpen
  \bibfield  {author} {\bibinfo {author} {\bibfnamefont {P.}~\bibnamefont
  {Nyhus}}, \bibinfo {author} {\bibfnamefont {S.~L.}\ \bibnamefont {Cooper}}, \
  and\ \bibinfo {author} {\bibfnamefont {Z.}~\bibnamefont {Fisk}},\ }\href
  {\doibase 10.1103/PhysRevB.51.15626} {\bibfield  {journal} {\bibinfo
  {journal} {Phys. Rev. B}\ }\textbf {\bibinfo {volume} {51}},\ \bibinfo
  {pages} {15626} (\bibinfo {year} {1995})}\BibitemShut {NoStop}%
\bibitem [{\citenamefont {Axe}\ and\ \citenamefont {Shirane}(1973)}]{Axe1973}%
  \BibitemOpen
  \bibfield  {author} {\bibinfo {author} {\bibfnamefont {J.~D.}\ \bibnamefont
  {Axe}}\ and\ \bibinfo {author} {\bibfnamefont {G.}~\bibnamefont {Shirane}},\
  }\href {\doibase 10.1103/PhysRevLett.30.214} {\bibfield  {journal} {\bibinfo
  {journal} {Phys. Rev. Lett.}\ }\textbf {\bibinfo {volume} {30}},\ \bibinfo
  {pages} {214} (\bibinfo {year} {1973})}\BibitemShut {NoStop}%
\bibitem [{\citenamefont {Ido}\ \emph {et~al.}(1990)\citenamefont {Ido},
  \citenamefont {Okayama}, \citenamefont {Ijiri},\ and\ \citenamefont
  {Okajima}}]{Ido1990}%
  \BibitemOpen
  \bibfield  {author} {\bibinfo {author} {\bibfnamefont {M.}~\bibnamefont
  {Ido}}, \bibinfo {author} {\bibfnamefont {Y.}~\bibnamefont {Okayama}},
  \bibinfo {author} {\bibfnamefont {T.}~\bibnamefont {Ijiri}}, \ and\ \bibinfo
  {author} {\bibfnamefont {Y.}~\bibnamefont {Okajima}},\ }\href {\doibase
  10.1143/JPSJ.59.1341} {\bibfield  {journal} {\bibinfo  {journal} {Journal of
  the Physical Society of Japan}\ }\textbf {\bibinfo {volume} {59}},\ \bibinfo
  {pages} {1341} (\bibinfo {year} {1990})}\BibitemShut {NoStop}%
\bibitem [{\citenamefont {Taniguti}\ \emph {et~al.}(2000)\citenamefont
  {Taniguti}, \citenamefont {Kuga}, \citenamefont {Okamoto}, \citenamefont
  {Kaneko},\ and\ \citenamefont {Ishihara}}]{Tan2000}%
  \BibitemOpen
  \bibfield  {author} {\bibinfo {author} {\bibfnamefont {H.}~\bibnamefont
  {Taniguti}}, \bibinfo {author} {\bibfnamefont {M.}~\bibnamefont {Kuga}},
  \bibinfo {author} {\bibfnamefont {H.}~\bibnamefont {Okamoto}}, \bibinfo
  {author} {\bibfnamefont {H.}~\bibnamefont {Kaneko}}, \ and\ \bibinfo {author}
  {\bibfnamefont {Y.}~\bibnamefont {Ishihara}},\ }\href {\doibase
  http://dx.doi.org/10.1016/S0921-4526(99)01240-5} {\bibfield  {journal}
  {\bibinfo  {journal} {Physica B: Condensed Matter}\ }\textbf {\bibinfo
  {volume} {291}},\ \bibinfo {pages} {135 } (\bibinfo {year}
  {2000})}\BibitemShut {NoStop}%
\bibitem [{\citenamefont {Yamaya}\ \emph {et~al.}(2012)\citenamefont {Yamaya},
  \citenamefont {Takayanagi},\ and\ \citenamefont {Tanda}}]{Yam2012}%
  \BibitemOpen
  \bibfield  {author} {\bibinfo {author} {\bibfnamefont {K.}~\bibnamefont
  {Yamaya}}, \bibinfo {author} {\bibfnamefont {S.}~\bibnamefont {Takayanagi}},
  \ and\ \bibinfo {author} {\bibfnamefont {S.}~\bibnamefont {Tanda}},\ }\href
  {\doibase 10.1103/PhysRevB.85.184513} {\bibfield  {journal} {\bibinfo
  {journal} {Phys. Rev. B}\ }\textbf {\bibinfo {volume} {85}},\ \bibinfo
  {pages} {184513} (\bibinfo {year} {2012})}\BibitemShut {NoStop}%
\bibitem [{\citenamefont {Zhu}\ \emph {et~al.}(2013)\citenamefont {Zhu},
  \citenamefont {Lv}, \citenamefont {Wei}, \citenamefont {Xue}, \citenamefont
  {Lorenz}, \citenamefont {Deng}, \citenamefont {Sun},\ and\ \citenamefont
  {Chu}}]{Zhu2013}%
  \BibitemOpen
  \bibfield  {author} {\bibinfo {author} {\bibfnamefont {X.}~\bibnamefont
  {Zhu}}, \bibinfo {author} {\bibfnamefont {B.}~\bibnamefont {Lv}}, \bibinfo
  {author} {\bibfnamefont {F.}~\bibnamefont {Wei}}, \bibinfo {author}
  {\bibfnamefont {Y.}~\bibnamefont {Xue}}, \bibinfo {author} {\bibfnamefont
  {B.}~\bibnamefont {Lorenz}}, \bibinfo {author} {\bibfnamefont
  {L.}~\bibnamefont {Deng}}, \bibinfo {author} {\bibfnamefont {Y.}~\bibnamefont
  {Sun}}, \ and\ \bibinfo {author} {\bibfnamefont {C.-W.}\ \bibnamefont
  {Chu}},\ }\href {\doibase 10.1103/PhysRevB.87.024508} {\bibfield  {journal}
  {\bibinfo  {journal} {Phys. Rev. B}\ }\textbf {\bibinfo {volume} {87}},\
  \bibinfo {pages} {024508} (\bibinfo {year} {2013})}\BibitemShut {NoStop}%
\end{thebibliography}
\end{document}